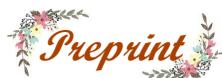
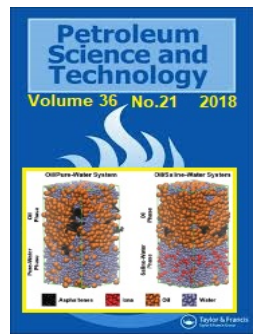

# Asphaltene aggregation onset during high-salinity waterflooding of reservoirs (a molecular dynamic study)

Salah Yaseen[a] and G.Ali Mansoori[b]

[a]Dep't of Chemical Engineering, Univ. of Illinois at Chicago, Chicago, Illinois, USA; *syasee3@uic.edu*
[b]Dep'ts of Bio- & Chemical Engineering, Univ. of Illinois at Chicago, Chicago, Illinois, USA; *mansoori@uic.edu*

**ABSTRACT**

The primary objective of this study is to establish an understanding of the role of high-salinity brine on the intensity of asphaltene aggregation onset during waterflooding of petroleum reservoirs. We already have shown that asphaltenes have a high tendency to form aggregates during waterflooding process when pure- and low salinity-water are injected into reservoirs. To fulfill the present objective, molecular dynamic simulations are per-formed on asphaltenic-oil/aqueous systems at 550 K-200 bar. The oil phase consists of asphaltenes (10 wt.%) and ortho-xylene, in which asphaltene molecules are completely soluble. Our simulations results reveal that the "salt-in effect" of high-salinity brine (25 wt.% NaCl) on seven different model asphaltenic oils causes a significant reduction of the onset of asphaltene aggregation as compared with pure-water. Such "salt-in effect" is primarily due to a considerable reduction of water miscibility in the oil phase at high pressure and temperature.

## 1. Introduction

Asphaltenes are a solubility class of compounds in crude oils. They are insoluble in paraffinic media, such as n-heptane, while they are soluble in aromatic crudes and solvents, such as ortho-xylene (Priyanto, Mansoori, and Suwono 2001; Mansoori 2009). It has been suggested that asphaltenes tend to form aggregates and deposit on surfaces. Deposited asphaltenes decrease production rate of crude oils by blocking the flow paths (Escobedo and Mansoori 1997; Vazquez and Mansoori 2000); Branco et al. 2001; Hu et al. 2004).

During petroleum production from reservoirs, crude oil is usually co-produced with substantial amounts of saline-water. The source of such saline-water is either: (*i*). The water that may exist in the reservoir rock pores within the hydrocarbon zone; and (*ii*). The water that is injected into the reservoir during waterflooding processes (Kim, Boudh-Hir, and Mansoori 1990; Pacheco-Sanchez and Mansoori 1997). There is a lack of understanding about the impact of saline-water on asphaltene aggregation and deposition from a petroleum reservoir during waterflooding, in reservoir conditions, at high pressures and temperatures, which motivated us to undertake the present study.

Molecular dynamics (MD) simulation has been applied with much success to investigate the behavior of asphaltenes in various model petroleum fluids (Pacheco-Sánchez, Zaragoza, and Martínez-Magadán 2004; Takanohashi, Sato, and Tanaka 2004; Hu et al. 2011; Yaseen and

**Corresponding author:** Salah Yaseen; *syasee3@uic.edu*; *salahyaseen1983@gmail.com*

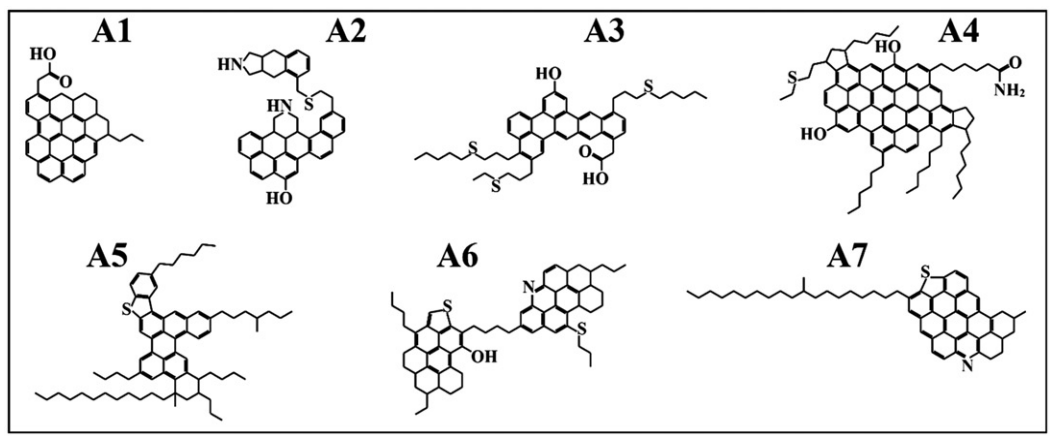

Figure 1. Molecular structures of the seven model-asphaltenes employed in this study.

Mansoori 2017; Khalaf and Mansoori 2018; Mohammed and Mansoori 2018). Recently, we investigated the onset of asphaltene aggregation during waterflooding process. Our MD simulations were performed on asphaltenic-oil/pure-water system at high temperature and pressure (550 K – 200 bar). MD results showed that asphaltenes became poorly soluble when water molecules were miscibilized in the oil phase. When the low-salinity brine (3.5 wt.% NaCl) was used instead of pure-water, the intensity of asphaltene aggregation onset dropped slightly due to the salt-in effect (Yaseen and Mansoori 2018).

The goal of the current report is to understand asphaltene stability when a high-salinity brine is injected into an asphaltenic-oil under reservoir conditions. We perform a series of MD simulations to investigate the intensity of asphaltene aggregation in well-defined oil and aqueous systems. Ortho-xylene is selected to be the oil medium since it is the best hydrocarbon solvent for asphaltenes. The onset measurements reported here should be valid for asphaltene molecules in crude oils (Yaseen and Mansoori, 2018).

## 2. MD simulation methodology

We perform MD simulations using GROMACS (version 5.1.2) simulation package (www.gromacs.org). In every MD simulation, we use OPLS-AA force field to represent asphaltenes, ortho-xylene, and salt-ions. Moreover, we employ SPC/E potential (Kusalik and Svishchev. 1994) to model water molecules. We use NPT ensemble to perform all MD simulations at constant temperature (550 K) and pressure (200 bar) for 20 ns. The detailed description of the MD method can be found in our previous publications (Yaseen and Mansoori 2017, 2018).

The simulation box is constructed such that it is composed of two phases: the aqueous phase and oil phase. The aqueous phase is either pure- or saline-water. The oil phase is always selected to be composed of asphaltene (10 wt.%) and ortho-xylene. Seven model-asphaltenes identified as A1, A2, …, A7, are employed in this report. The molecular structures of these model-asphaltenes are depicted in Figure 1. The number of asphaltene molecules in every simulation box is maintained constant at 24.

To determine the effect of the water salinity on asphaltene stability, three groups of MD simulations are performed. Each group is composed of seven MD simulations. In the first group of MD simulations, which is termed as PW, we use pure-water. In the second group of MD simulations, which is termed as SW1, we use saline-water (25 wt.% NaCl). In the third group of MD simulations, which is termed as SW2, we use saline-water (25 wt.% natural-salt) that has an ionic composition like that of ocean water. The molar composition of SW2 salt-ions is: calcium (2.5%), chlorine (47%), potassium (2%), magnesium (6%), sodium (39%), sulfate (3.5%).

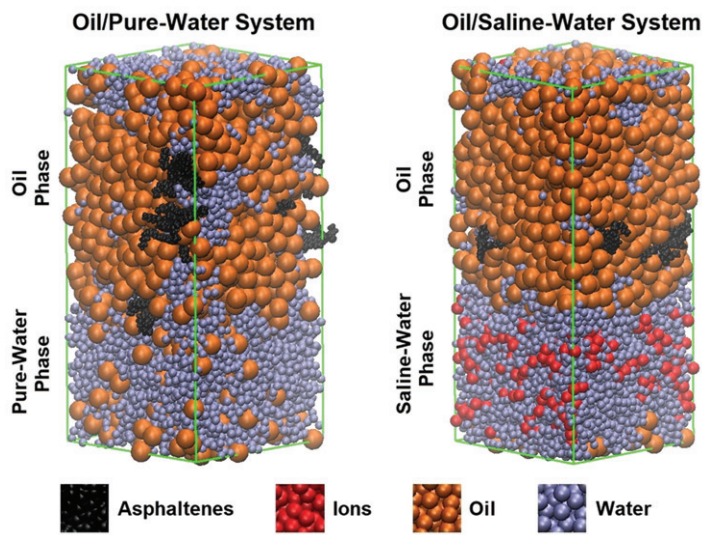

**Figure 2.** Schematic representation of the oil/pure-water and oil/saline-water systems at 550 K and 200 bar.

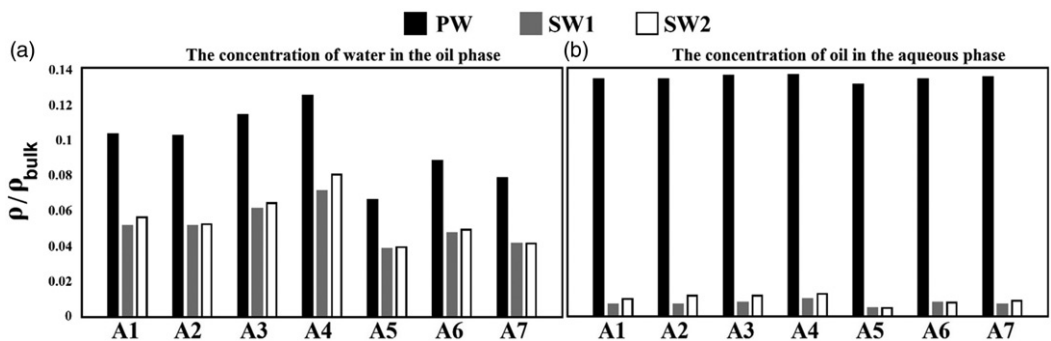

**Figure 3.** The concentration of water and oil in oil/aqueous systems. MD calculations are performed at 550 K and 200 bar. In this figure and the other figures $\rho$ stands for density, PW stands for pure-water, SW1 stands for 25 wt.% NaCl-brine, and SW2 stands for 25 wt.% natural-salt-brine.

## 3. Results

### 3.1. Structural analysis of oil/aqueous systems

The structural analysis of MD systems is investigated using Figure 2, which shows a schematic representation of the oil/aqueous systems. It is obvious that asphaltene molecules stay suspended in the oil phase due to the strong van der Waals interaction between asphaltenes and oil (Yaseen and Mansoori 2017). Similarly, salt-ions do not diffuse into the oil phase due to the strong interactions between water and ions (Yaseen and Mansoori 2018). On the other hand, the schematic representation shows that oil and water present mutual miscibility. According to Figure 3, The concentration of water in the oil phase is ranged between 4 and 14 wt.%. Furthermore, the concentration of oil in the aqueous phase is ranged between 1 and 14 wt.%.

### 3.2. The radial distribution functions (RDFs) of asphaltenes

RDFs of the center of mass of asphaltene moieties are calculated (see Figure 4). The heights of RDF peaks are used to compare the intensity of asphaltene aggregation of various simulations (Mohammed and Mansoori 2018; Yaseen and Mansoori 2018). PW exhibit the highest RDF peaks, which indicate that pure-water causes the highest intensity of asphaltene aggregation. Interestingly, asphaltene aggregation intensity is reduced sharply when the high-salinity brines are used (salt-in effect). It decreases by (65%–91%) and (59%–88%) during SW1 and SW2, respectively. In addition, RDFs illustrate that the high-salinity NaCl-brine is more effective in asphaltene solubilization than that of the natural-salt.

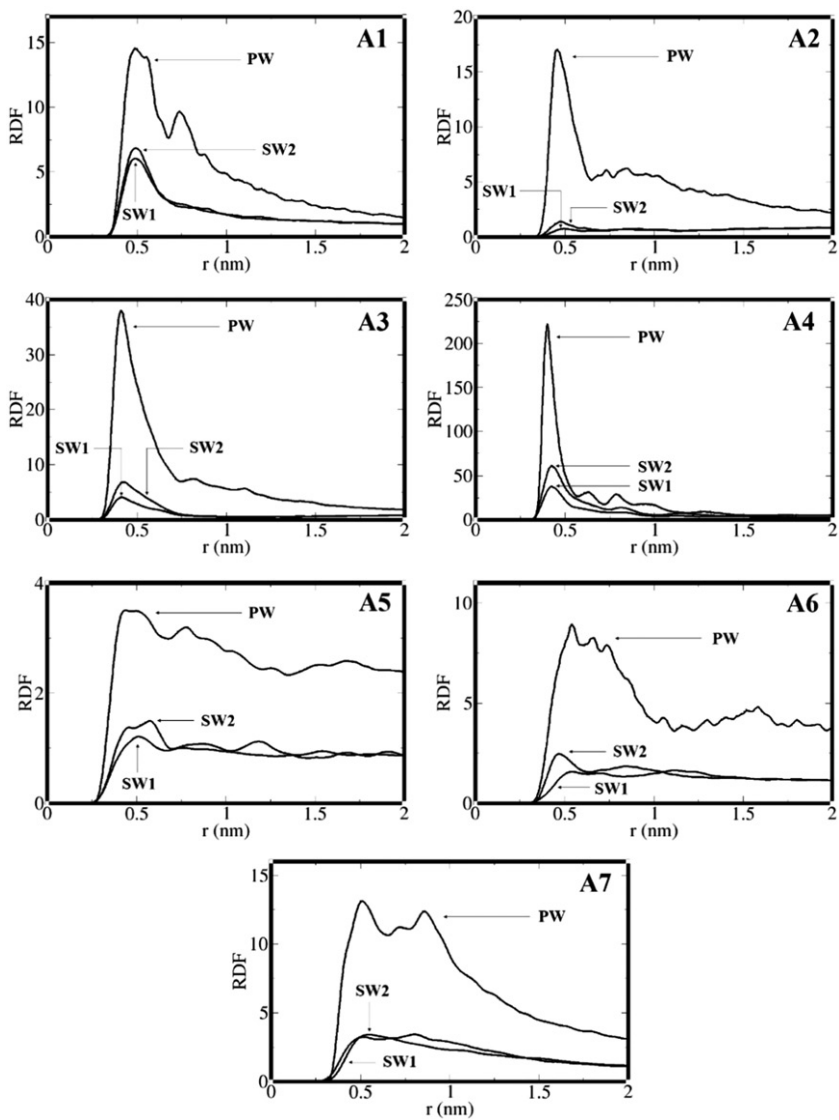

**Figure 4.** The RDFs of the center of mass of asphaltene moieties. MD calculations are performed at 550 K and 200 bar.

Furthermore, we use RDFs to shed light on the effect of high-salinity-brines on the configuration of stacked asphaltenes. RDFs of every model-asphaltene show somewhat similar shapes during PW, SW1, and SW2. RDF curves exhibit high peaks at separation distances within (0.38 nm–0.50 nm), which corresponds to neighboring molecule of the same in the stacking structure of asphaltenes. Such separation distances indicate that the predominant structural contact between the associated asphaltene molecules is face-to-face stacking as we reported before (Yaseen and Mansoori 2018). This suggests that the high-salinity brines do not greatly affect the stacking configuration of the aggregated asphaltene molecules.

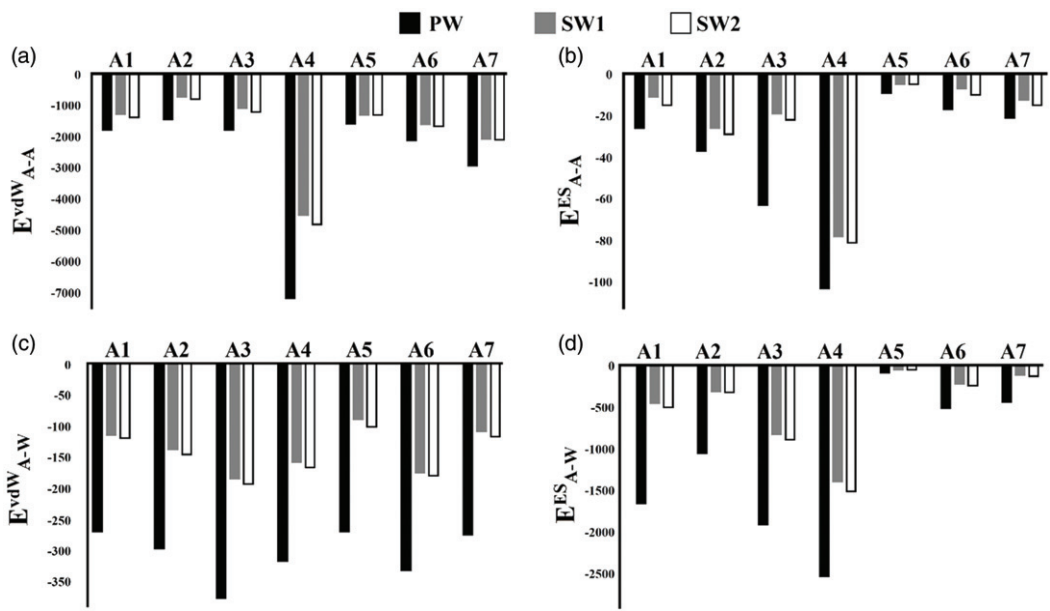

**Figure 5.** Potential energies [kJ/mol] of asphaltene-asphaltene ($E^{vdW}_{A-A}$ and $E^{ES}_{A-A}$) and asphaltene-water ($E^{vdW}_{A-W}$ and $E^{ES}_{A-W}$). MD calculations are performed at 550 K and 200 bar.

### 3.3. Potential energies of asphaltenes

We also investigate the molecular interactions between asphaltene molecules by computing electrostatic ($E^{ES}_{A-A}$) and van der Waals ($E^{vdW}_{A-A}$) potential energies (see Figure 5a and 5b). In here, the calculation of the potential energies is used to compare the intensity of asphaltene aggregation. The aggregated asphaltenes always have lower potential energies than individual dissolved ones.

Results indicate that van der Waals forces are the prevailing interaction between asphaltenes. This finding is in line with many previous MD studies that reported asphaltene aggregation (see for example: Khalaf and Mansoori, 2018; Sedghi et al., 2013; Silva et al., 2016). PW group exhibits the greatest attraction between asphaltene molecules. Attraction energies significantly decrease when the high-salinity brines are used instead of pure-water. The lowest attraction energies between asphaltene molecules are observed when high-salinity NaCl is used for SW1 group.

### 3.4. Distribution of asphaltene aggregates

We further assess the impact of water salinity on asphaltene association by investigating the distribution of asphaltene aggregates as reported in Figure 6. The distribution of asphaltenes is reported here using the average fraction of asphaltene in various association stages (monomer, dimer, trimer, etc.).

The fraction of asphaltene molecules that form aggregates is diverse, and it depends on the type of model-asphaltene and water salinity.

These results suggest that the injection of high-salinity-brines significantly solubilize all model-asphaltenes except A4 by increasing the fraction of asphaltene monomers. In the case of A4, it is obvious that high-salinity-brines reduce slightly the fraction of asphaltene molecules that form aggregates, which may be attributed to the large aromatic moiety of A4 that is composed of 16 aromatic rings. However, their large aggregates (heptadecamers in PW) break into smaller ones (dimers, trimers, tetramers, pentamers, and hexamers in SW1 and SW2).

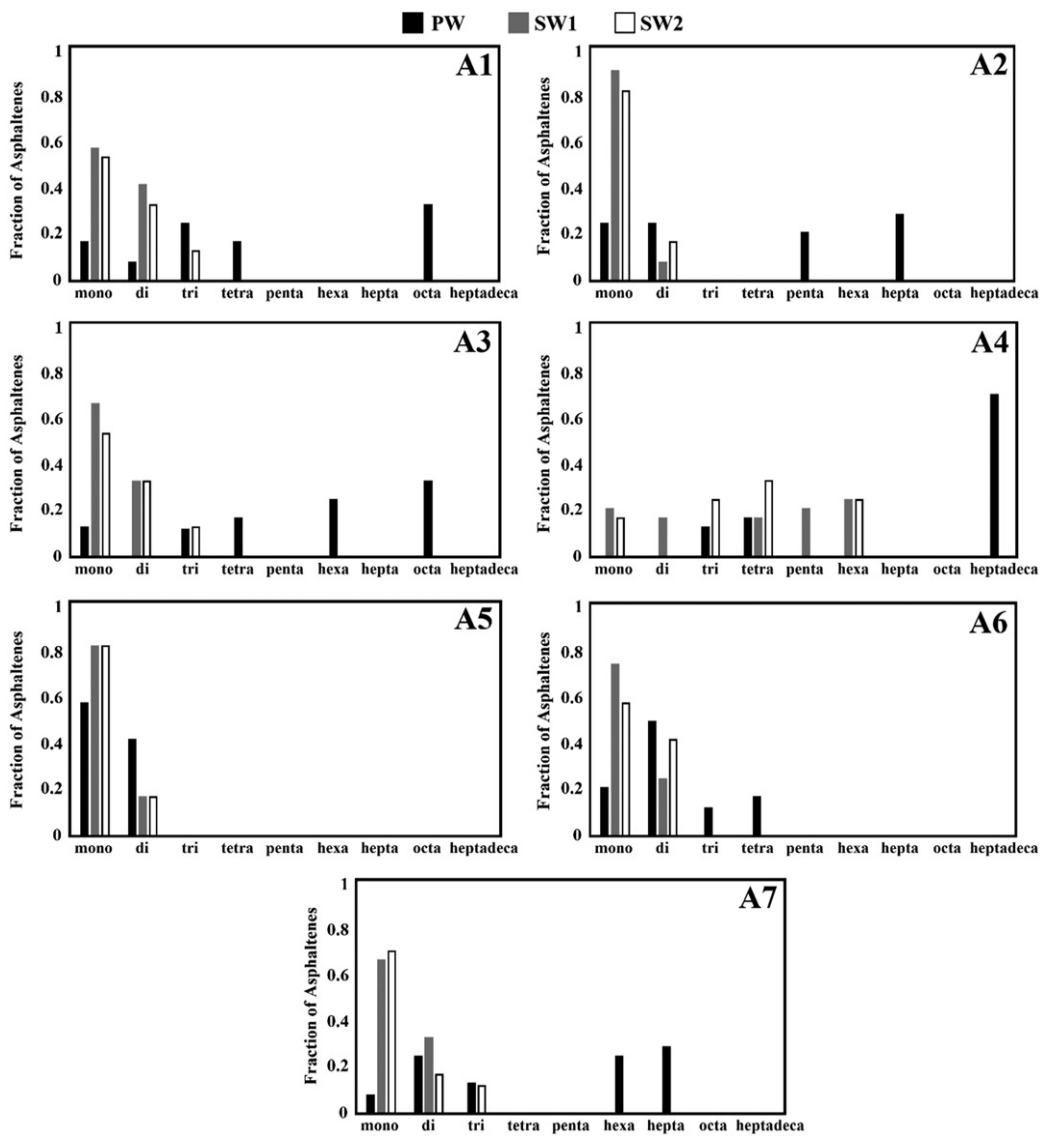

**Figure 6.** The average fraction of asphaltene in various association stages. MD calculations are performed at 550 K and 200 bar. In this figure mono stands for monomer, di stands for dimer, tri stands for trimer, …., heptadeca stands for heptadecamer.

## 4. Discussion

RDFs, potential energies, and the distribution of asphaltene molecules in monomers and various aggregate sizes show the salt-in effect on asphaltenes when high-salinity brines are injected into asphaltenic-oil under the high temperature and pressure of (550 K and 200 bar) representing an extreme reservoir condition. Additionally, they demonstrate that NaCl-brine is more effective in asphaltene solubilization than natural-salt-brine. Besides, our results indicate that using high-salinity-brines does not affect the stacking configuration of asphaltene molecules in their aggregates.

Figure 3 illustrates that the variation of water salinity affects the concentration of water in the oil phase. The highest concentration of miscibilized water is observed during PW. The miscibilized water concentration decreases sharply by 36-50% when we use high-salinity brines instead of pure-water. In a previous publication we already have demonstrated that miscibilized water is the main cause for asphaltene association in oil/aqueous system (Yaseen and Mansoori 2018). The depression of the miscibilized water concentration reduces to a large extent the interactions between asphaltene and water molecules (see Figure 5c and 5d). That decreases the population of the miscibilized water near asphaltene molecules and diminish their ability to destabilize asphaltenes.

Figure 3 also demonstrates that the miscibility of oil in the aqueous phase is influenced by the variation of water salinity. The concentration of the miscibilized oil is highly reduced by (86%–93%) when the high-salinity brines are employed instead of pure-water. Consequently, the concentration of ortho-xylene in the oil phase is increased. This enhances $\pi$-$\pi$ interaction between ortho-xylene and asphaltene molecules and decreases the intensity of asphaltene aggregation.

## 5. Conclusions

The "Salt-in effect" is found to be the dominant effect on the asphaltene aggregation onset when a high-salinity brine is injected into asphaltenic-oil during waterflooding. The intensity of asphaltene aggregation onset is remarkably reduced by about 60–90% when a pure-water is replaced by a high-salinity brine (25 wt.% NaCl). Moreover, the NaCl-brine is more effective in asphaltene solubilization than that of natural-salt.

We conclude that the salt-in of asphaltene is mainly due to the considerable reduction of water miscibility in the oil phase. The water concentration in the oil phase drops markedly by an average of 43% when pure-water is replaced by high-salinity brine (25 wt.% NaCl).

We are aware of the fact that MD simulation is incapable of predicting the behavior of complex petroleum fluids. What we have reported here is the onset of interaction/aggregation of two individual dissolved-in-oil asphaltene molecules as a result of the presence of miscible water in oil. We are confident this result is applicable to asphaltenes behavior in real crude oils.

**Acknowledgments**: Salah Yaseen is grateful to the Higher Committee for Education Development in Iraq (HCED) for his financial support during this research.

## ORCID

Salah Yaseen  https://orcid.org/0000-0003-0724-8044
G.Ali Mansoori  https://orcid.org/0000-0003-3497-8720

## Nomenclature

| | |
|---|---|
| Ai | i = 1, 2, …, 7, notations for model-asphaltenes used in this study |
| Di | Dimer |
| $E^{vdW}_{A-A}$ | Van der Waals potential energy between asphaltenes |
| $E^{vdW}_{A-W}$ | Van der Waals potential energy between asphaltenes and water |
| $E^{ES}_{A-A}$ | Electrostatic potential energy between asphaltenes |
| $E^{ES}_{A-W}$ | Electrostatic potential energy between asphaltenes and water |
| Hepta | Heptamer |
| heptadeca | Heptadecamer |
| Hexa | Hexamer |
| MD | Molecular dynamics |
| Mono | Monomer |
| Octa | Octamer |
| Penta | Pentamer |
| PW | Group of simulations that use pure-water |
| RDF | Radial distribution function |
| SW1 | Group of simulations that use 25 wt.% NaCl |
| SW2 | Group of simulations that use 25 wt.% natural-salt |
| Tetra | Tetramer |
| Tri | Trimer |
| $\rho$ | Density |